%% file: main.tex
\documentclass[conference, 12pt]{IEEEtran}

\usepackage{graphicx, subfigure}
\usepackage{amsmath, amssymb}
\usepackage{listings}
\usepackage{verbatim}
\usepackage{zlmtt}
\usepackage{fancyvrb}
\usepackage{subfiles}
\usepackage{enumitem}
\usepackage{color, colortbl}
\usepackage{tikz}
\usepackage{caption} 
\usepackage{nicematrix}
\usepackage{soul}
\usepackage{hyperref}
\definecolor{Gray}{gray}{0.9}

\AtBeginDocument{%
  \providecommand\BibTeX{{%
    \normalfont B\kern-0.5em{\scshape i\kern-0.25em b}\kern-0.8em\TeX}}}

\begin{document}

\title{ClusterLog: Clustering Logs for Effective Log-based Anomaly Detection}

\author{\IEEEauthorblockN{Chris Egersdoerfer, Di Zhang}
\IEEEauthorblockA{\textit{Computer Science Department} \\
\textit{University of North Carolina at Charlotte}\\
Charlotte, NC, United States \\
cegersdo@uncc.edu, dzhang16@uncc.edu}

\and
\IEEEauthorblockN{Dong Dai}
\IEEEauthorblockA{\textit{Computer Science Department} \\
\textit{University of North Carolina at Charlotte}\\
Charlotte, NC, United States \\
ddai@uncc.edu}}

\maketitle

%
\begin{abstract}
With the increasing prevalence of scalable file systems in the context of High Performance Computing (HPC), the importance of accurate anomaly detection on runtime logs is increasing. But as it currently stands, many state-of-the-art methods for log-based anomaly detection, such as DeepLog, have encountered numerous challenges when applied to logs from many parallel file systems (PFSes), often due to their irregularity and ambiguity in time-based log sequences.
To circumvent these problems, this study proposes ClusterLog, a log pre-processing method that clusters the temporal sequence of log keys based on their semantic similarity. By grouping semantically and sentimentally similar logs, this approach aims to represent log sequences with the smallest amount of unique log keys, intending to improve the ability for a downstream sequence-based model to effectively learn the log patterns. The preliminary results of ClusterLog indicate not only its effectiveness in reducing the granularity of log sequences without the loss of important sequence information but also its generalizability to different file systems’ logs.
\end{abstract}
\section{Introduction}
\subfile{Intro}

\section{Design and Implementation}
\label{design}
\subfile{DesignAndImplement}

\section{Evaluation}
\label{evaluation}
\subfile{Evaluation}

\section{Related Works}
\label{related}
\subfile{RelatedWork}

\section{Conclusion and Future Work}
\label{conclusion}
\subfile{Conclusion}

\bibliographystyle{plain}
\bibliography{bib}

\section*{Acknowledgements}
We sincerely thank the anonymous reviewers for their valuable feedback. This work is partially supported by NSF grants CCF-1908843, CCF-1910727, and CNS-2008265.

\end{document}

%% file: Intro.tex
In light of growing datasets, increasing problem complexity, and more compute intensive algorithms, it is no question that large-scale computing systems, such as Cloud or High-Performance Computing (HPC), are a growing area of interest. In these systems, distributed storage frameworks are the critical foundation to providing global data access, and their health is therefore critical to the function of the entire system. However, due to increasing scale and complexity, distributed storage systems are subject to various bugs, failures, and anomalies in production, which lead to data loss, service outages and degradation of quality of service~\cite{gupta2017failures, das2018doomsday, pfault}. It is thereby critical to detect malfunctions accurately and in a timely manner, so that system operators can promptly pinpoint issues and resolve them immediately to mitigate losses.

It has been proven that runtime logs, which record detailed runtime information about storage systems during operation, are a valuable information source to detect potential anomalies in distributed storage systems. These runtime logs are generated by log statements written in the source code (using simple \texttt{printf} function or logging libraries such as Log4j~\cite{log4j}). These logs record important internal states of distributed storage systems, such as key variable values, return values, and performance statistics, all of which can be useful to reveal system anomalies. As a result, an extensive amount of research on log-based anomaly detection has been done recently~\cite{cinque2012event,roy2015perfaugur, hansen1993automated,oprea2015detection,yamanishi2005dynamic,debnath2018loglens,chen2004failure,liang2007failure,du2017deeplog,meng2019loganomaly,xu2009online, xu2009detecting,lou2010mining,lin2016log,das2018doomsday}.

The main theme of modern Log-based anomaly detection solutions is to apply machine learning, especially deep learning methods, onto log sequences to detect anomalies. The common process of doing so is for logs to be parsed and processed first, then their normal sequence to be learned by sequence-based pattern detection models such as LSTM~\cite{le2021log}. Having learned what normal and abnormal log sequences look like, these algorithms are then shown runtime logs and are tasked to classify them accurately. Take one of the most representative works, DeepLog~\cite{du2017deeplog}, as an example: during training, it first parses the runtime logs into templates and represents each of these templates using a single integer. Through this process, a sequence of logs becomes a sequence of integral identifiers, which will be learned using an LSTM model. During runtime, the anomalies are defined by whether or not the actual next log identifier is within the set of identifiers which the LSTM model predicts. In another recent study, NeuraLog~\cite{le2021log}, the runtime logs are identified as a vector instead of an integer to contain more semantic information. Still, these sequences of embedded vectors will are fed into a DL model to learn the normal/abnormal sequences.

Although these sequence-based log anomaly detection solutions work well for many storage systems such as HDFS~\cite{hdfs}, they have one key issue: they rely heavily on the quality of the log sequences in the training data. The sequence of logs must be both accurate and representative of the log system's logic. Such sequences of logs are often expensive to obtain in the real-world. 
For instance, the HDFS logs used in existing studies were pre-processed by aggregating logs with the same data block ID into a sequence, regardless of how far these log entries are from each other in the runtime. Unfortunately, such pre-processing is not always possible. In fact, many parallel storage systems, such as Lustre~\cite{lustre} and BeeGFS~\cite{beegfs}, do not have any common identifier (ID) in log entries to denote their relevance. Missing such global IDs makes it difficult to identify the matching events or to build log sequences accurately, resulting in only raw time-based log sequences available~\cite{zhang2021sentilog}.

In addition, the raw log sequences generated from a distributed environment are quite ambiguous. One source of ambiguity is the \textit{clock skew} in distributed systems, as logs generated concurrently across multiple nodes are merged into a single file where their order is not always equivalent to the order of time in which they occurred. Secondly, interleaved concurrent threads present a further, and more complex problem. As different nodes run separate execution threads concurrently, the unrelated processes are often logged in random, interleaving order. Directly learning from these often noisy sequences of runtime logs, can be problematic, and require a much larger labeled dataset and longer training time.


To address these issues, in this study we propose ClusterLog, a {log pre-processing} method which clusters individual runtime logs based on their similarity to effectively reduce the ambiguity of log sequences and improve the accuracy of downstream sequence-based learning. 
The intuition behind ClusterLog is driven by the idea that grouping similar runtime logs together will result in less random variation within the log sequence due to a lesser amount of unique key identifiers. In addition, grouping logs based on their similarity can still retain the vital sequence information between different types of high-level file system operations. For example, Lustre log sequences contain many sequences of logs where actions are all very similar, but because of the lack of block ID, they are highly irregular in time sequence. Grouping some of these similar actions is intended to eliminate a large portion of this irregularity, providing a cleaner sequence to be learned. Further, the robust and generalizable nature of this approach allows it to be applied to numerous types of file system logs and on limited amounts of available training data, both of which are not adequately captured by previous approaches. 

The rest of this paper is organized as follows. In Section~\ref{design}, we present the design and implementation of ClusterLog. In Section~\ref{evaluation}, we discuss the evaluation setup and results. Finally, in sections ~\ref{related} and \ref{conclusion}, we discuss related work and lay out our future work, respectively.

%% file: DesignAndImplement.tex
The implementation of ClusterLog can be effectively broken into four parts. The first is rudimentary preprocessing, {\color{black} where the log content is extracted from the log files}, resulting in only the natural language sequence of each log which can be matched throughout the log file to create a set of unique log keys. From here, the preprocessed log keys are fed into a pre-trained semantic similarity model to produce unique embeddings for each unique log key. Simultaneously (or in sequence) to the semantic similarity embedding step, the preprocessed log keys are fed into a pre-trained sentiment prediction model to result in a 0 or 1 prediction of each log’s sentiment. The output of the semantic similarity embedding model and the sentiment prediction model are concatenated at this point and serve as the entire latent representation of each log key. Following the concatenation, the embeddings are fit to a clustering algorithm where the resulting cluster labels are used to replace the original sequence of log keys. Finally, the sequence of these cluster labels are fed into the downstream sequential analysis algorithm. In our current implementation, we use DeepLog's sequence-learning part as the downstream algorithm. It is a two-layer LSTM network which is used to predict a probability distribution of the next log key given a specified window of previous logs. If the next log key is within the top candidates of this distribution, the sequence is considered normal, otherwise it is labeled as anomalous. The code base which implements the design described in this section is available using the following link: \url{https://github.com/DIR-LAB/ClusterLog}.

\subsection{Preprocessing}
The first step of the ClusterLog implementation is rudimentary preprocessing of the raw log files. The goal of this step is to remove any excess characters and information from each individual log key, to result in a stripped key containing only the natural language sequence of the log which can be extracted as a unique log key and matched with many other logs in the original file. While it is ideal to reduce the log to its most generic natural language form to result in the smallest possible amount of unique log keys, an approximation of this will likely suffice if the former is too difficult to achieve. This is sufficient as the semantic embedding model will almost identically embed keys which are of the same logging template but are not a perfect match, meaning they will be clustered together even at the lowest of distance thresholds. As preprocessing can occasionally be very tedious, approximation at this step is highly valuable in regards to time and effort required to set up this approach compared to others. Depending on the file system which is being analyzed, the exact preprocessing steps may vary, but with the same goal in mind, preprocessing generally consists of extracting the log content, removing unique identifiers such as time stamps or Block IDs, removing unnecessary characters or character patterns, and unabbreviating words.

\subsection{Semantic Embedding}
Once the logs have been preprocessed, the natural language sequence of each log is fed into a pre-trained semantic embedding model. Though a number of applicable models do exist, the most accurate embeddings were achieved using the all-mpnet-base-v2 \cite{allMPnet2} model which itself was pre-trained on the original mpnet-base \cite{SongMPnet} model, {\color{black}published in 2020}, and further fine-tuned on over 2.1 Billion sentence pairs to optimize the model for semantic similarity. Though a further fine-tuning step specific to the semantic similarity among log keys would likely improve the semantic embedding quality even further, there is no known labeled dataset for this task, and creating labeled sentence pair data for semantic similarity is an arduous task. The output of this embedding model is a latent representation of the natural language contained within the log key of shape (768, 1) {\color{black}where the distance in the latent space between semantically similar sentences is close together, and those which are semantically diverse are far apart.}

\subsection{Sentiment Prediction}
While the semantic embedding model is valuable in that it creates a latent space where semantically (language-wise) similar sentences are close together and diverse ones are far apart, which can be used to cluster log keys reporting on the same or similar processes, these embeddings sometimes miss a valuable feature of natural language which is often included in logs. That feature is sentiment. A simple example of how classifying sentiment can add value to a semantic embedding is evident in the following two HDFS logs: \texttt{Exception in receiveBlock for block} and  \texttt{Receiving empty packet for block}. These two logs are similar in terms of semantics using our pre-trained model because of their shared key words. However, the first one indicates an anomaly while the second does not. 
This presents a challenge when clustering solely based on semantic embedding. 

To work around this kind of issue, we 
follow the idea of our recent SentiLog work which leverages the sentiment of file system logs~\cite{zhang2021sentilog}. Specifically, we reuse 
the pre-trained sentiment classification model from SentiLog on a set of neutral/positive and negative log keys, and concatenate the rounded (0 or 1) output of this model to the overall embedding of the log. 
Adding a sentiment dimension properly helps separate logs which may be semantically similar but opposite with regard to sentiment. Ultimately, the semantic embedding, including the concatenated sentiment prediction, serves as a highly accurate latent representation of the log key which can confidently be used to cluster logs which are truly similar. 

\subsection{Clustering}
The ultimate goal of ClusterLog is to cluster the runtime logs into similar groups, so that we can use the same group ID to represent similar logs to reduce the complexity of the log sequences. 

Following the semantic embedding and sentiment prediction steps, and their concatenation, the entire embedding is fed into a clustering algorithm in order to group the keys. 
However, there are a variety of problems associated with common clustering algorithms such as K-Means clustering when applied to this task. Initially, we ruled out K-Means primarily because of the need to specify the amount of centroids, which presents a major challenge as it makes the approach highly dependent upon the training data being fully representative of the logs during deployment as new, unseen, log keys would be forced to be grouped with an existing cluster, regardless of actual semantic and sentiment similarity. To add to this, finding an optimal number of centroids in K-Means based purely on the embedding data presents a challenge in and of itself as classic methods like the Elbow method are inconclusive on out our datasets.


In response to this, other clustering methodologies which create an arbitrary amount of clusters dependent on a specific distance threshold between clusters or points seem to be more suited. However, in practice, these too can be difficult to assess, as finding the correct hyper-parameters (i.e. the distance) for the given dataset is not always clear. But through extensive exploration, Density-Based Spatial Clustering of Applications with Noise (DBSCAN) \cite{DengDBSCAN} gave the most promising results when applied to a variety of file systems’ logs. DBSCAN is simply described as a clustering algorithm which defines data points as core points, border points, and noise points based on \textit{Epsilon}, a distance threshold, and \textit{minimum samples}, both of which must be manually set. 

\begin{figure}[h]
    \centering
    \includegraphics[width=0.7\linewidth]{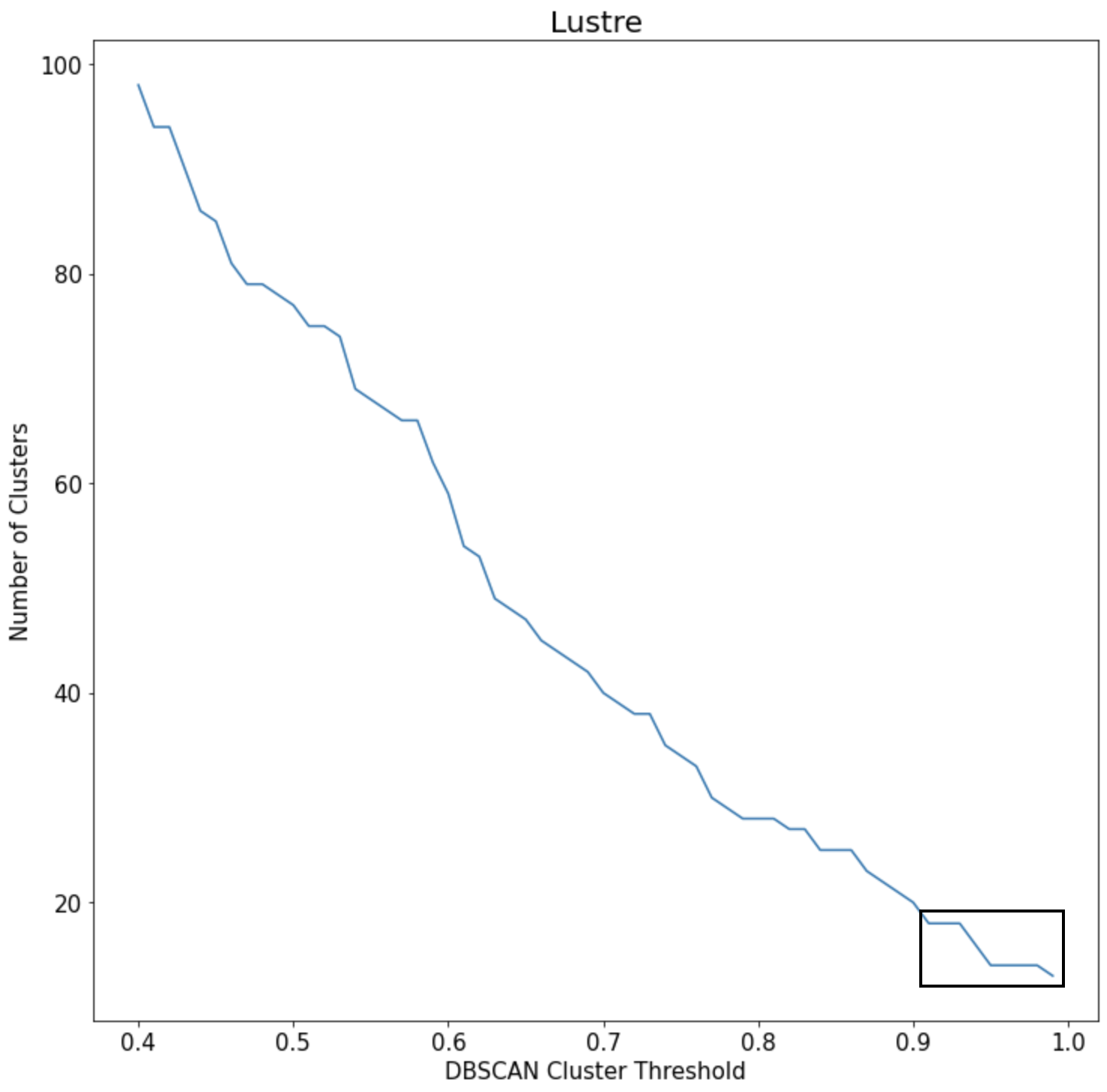}
    \caption{\small Number of generated clusters on \textit{Lustre} logs using different DBSCAN threshold.}
    \label{fig:LustreThresholds}
\end{figure}
\begin{figure}[h]
    \centering
    \includegraphics[width=0.7\linewidth]{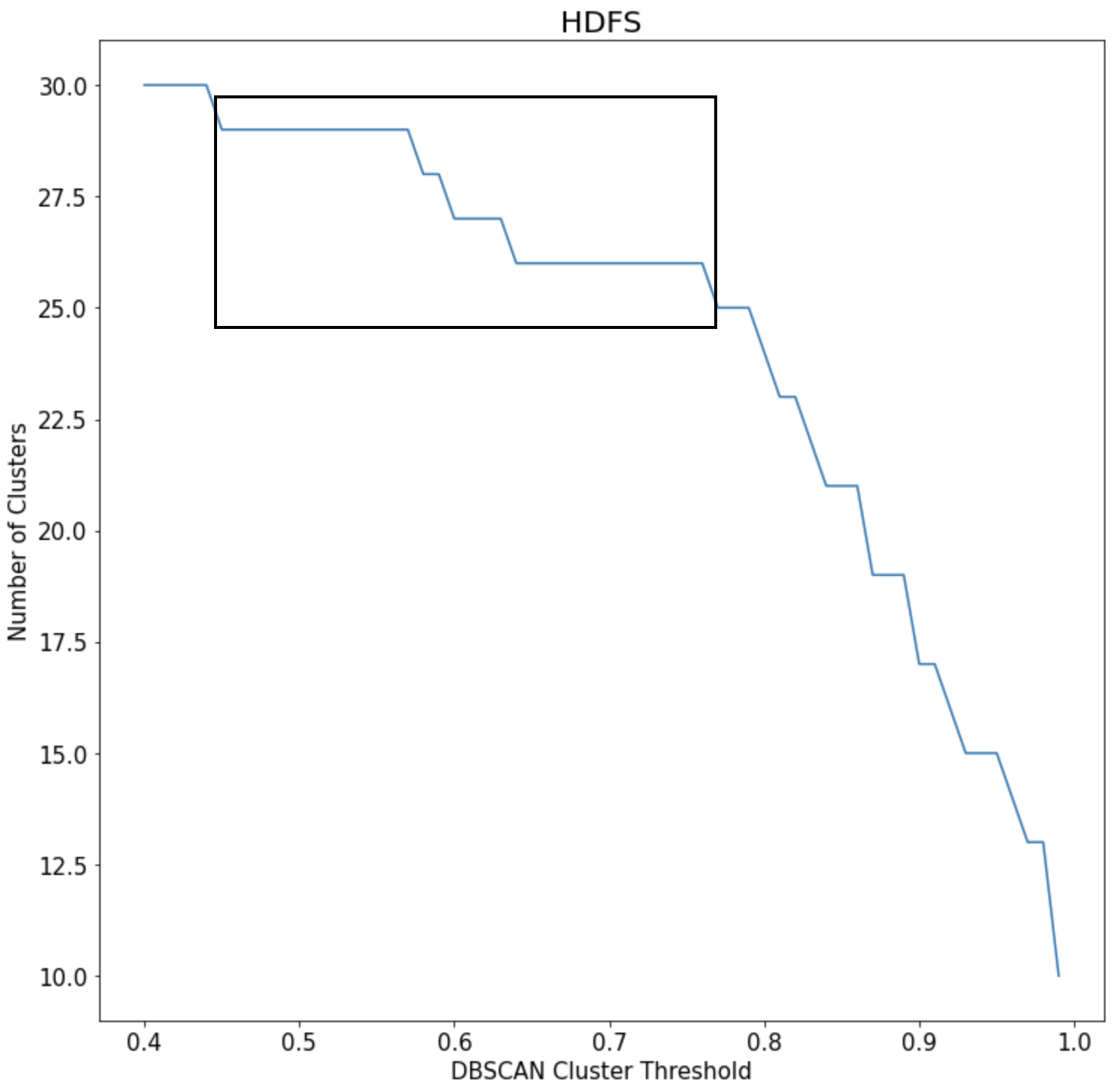}
    \caption{\small Number of generated clusters on \textit{HDFS} logs using different DBSCAN threshold.}
    \label{fig:HDFSThresholds}
\end{figure}
The reason for choosing DBSCAN was primarily driven by the fact that it gave a more consistent, and often more valuable, insight on how to set the threshold hyper-parameter for good results. In contrast to KMeans and other classic algorithms, the epsilon parameter used by DBSCAN can be used to locate density gaps in the data as well as the amount of change required to overcome these gaps. Additionally, \textit{Epsilon} is highly intuitive to tune as it can be easily understood as modifying the radius around each point which is used to search for neighboring points. This means that when applied to any system, some analysis of how far apart embeddings of similar entries are can be used to gain meaningful insight as to what a good \textit{Epsilon} value may be. As shown in Figure \ref{fig:LustreThresholds} and \ref{fig:HDFSThresholds}, when plotting epsilon from 0.4 to 1 against the amount of clusters created by DBSCAN for HDFS and Lustre, respectively, both graphs indicate noticeable ledges where multiple steps of epsilon do not result in any change in the amount of clusters found by DBSCAN. The thresholds at or around the longest of these ledges, provide a good baseline for setting the epsilon hyperparameter. Using a very minimal training and testing set allows us to verify such hyperparameter values.

With regard to the second hyperparameter which was mentioned, \textit{minimum samples}, it was set to a constant value of 1. By setting minimum samples to 1, DBSCAN does not predict any of the data points as noise, but rather, any outliers simply become a unique cluster of just themselves. This was done for the simple reason that labeling all outlier points as -1 to indicate noise effectively creates a meaningless cluster of completely unrelated log keys which are only grouped because they are too far from others. This severely impacts the ability for the sequential analysis algorithm to properly predict anomalies.

\subsection{Sequential Analysis}
The final part of ClusterLog is the downstream sequential analysis algorithm. In the current implementation, we focus on comparing with a state-of-the-art log-based anomaly detection method, DeepLog. Hence, we use DeepLog's sequence-learning part as the downstream algorithm. It is based on a standard 2 layer LSTM architecture which uses ten sequential log keys as input to predict the probability distribution of the next log key. From this probability distribution, log keys within the top 9 are treated as part of a normal sequence, while other log keys in this distribution are treated as anomalies. 


%% file: Evaluation.tex
To evaluate the performance of ClusterLog, we conducted evaluations on two vastly different distributed storage systems: Apache HDFS and Lustre. 
The combination of these different evaluations demonstrates the generality of ClusterLog on both noisy Lustre logs and more structured HDFS logs. We will show that our approach will outperform existing solutions in both areas through granularity reduction.

\subsection{Lustre}
\subsubsection{Dataset}
Due to the lack of publicly available anomaly detection datasets for Parallel File Systems, the dataset utilized in this paper was generated and labeled via a process of fault injection. Specifically, the fault injection was simulated using PFault \cite{pfault}, an open-source fault injection repository for Parallel File Systems. The labeling of this data was done by treating data before fault injection as normal and having domain experts manually label the data following a fault injection to ensure its relevance with the injected anomaly. This process resulted in a dataset containing 150,473 normal logs and 7,401 abnormal, anomalous logs. Additionally, the total number of unique log key templates in this dataset equated to 73.

\subsubsection{Training and Testing}
The training setup for evaluating ClusterLog against the Lustre dataset can be simply described as learning a portion of the normal sequence of logs (without anomalies) to show the sequence model how normal logs look, and then using that knowledge to see if anomalies can be detected in a sequence of logs containing both normal and anomalous logs. Both the train and test set are created by forming a sliding window of size 10 among the log key sequence, where the goal is to create a probability distribution for the next log key. If the next log key is not within the top candidates of the probability distribution it is counted as an anomaly while testing. {\color{black}A distinction to this setup is made when comparing it with NeuralLog as NeuralLog trains on a set of both normal and anomalous logs. Additionally, NeuralLog does not predict the next log in a sequence, but rather classifies a given sequence as anomalous or not. Based on these approaches, Accuracy, Precision, Recall, and F-measure are calculated.}

While most of the training and testing analysis was done using 25 percent of the total normal logs, in order to test {and compare} the limits of generalizability, a further test was carried out in which a smaller amount of the entire dataset was used to train the sequence model. This test was carried out by {\color{black}using just 1 percent of the entire Lustre training set. This split resulted in ~1,500 total samples.}

\subsection{HDFS}
\subsubsection{Dataset}
Among the majority of recent anomaly detection works, the same HDFS dataset is most commonly referenced in their respective evaluations. This makes it a good benchmark comparison for new results. In contrast to logs contained in the Lustre log dataset, HDFS logs contain a specific block ID in addition to their log content which allows them to be grouped by session. Anomalies are labeled on a per-session basis rather than a per-log basis for this dataset. The labeling itself was carried out by domain experts who discovered and labeled sessions containing one of 11 types of anomalies. This resulted in a dataset with 11,197,954 messages, grouped into 575,139 sessions, of which 16,808 sessions were anomalous~\cite{xu2009online}.

\subsubsection{Training and Testing}
During training, the sequence model utilizes only the normal log sessions, and each key in each session is represented by the corresponding cluster calculated by DBSCAN. Much like what was done with Lustre, the sequence model learns normal behavior by using a window of 10 cluster IDs. During testing, the trained sequence model is run against both the normal and the anomalous log sessions. For each session, the sequence model utilizes the window of 10 keys to predicting a probability distribution for the next ID in the session. As opposed to Lustre, if the model predicts an anomaly, the session is labeled as an anomaly, instead of just the log key. If no anomalies are predicted within a given session, the session itself is predicted as being normal. {\color{black}This setup varied slightly in the case of Neuralog, as it did for Lustre, as Neuralog views each entire block and classifies the block as anomalous or normal.}


\subsection{Results Analysis}
\begin{figure*}[t]
    \centering
    \includegraphics[width=1\linewidth]{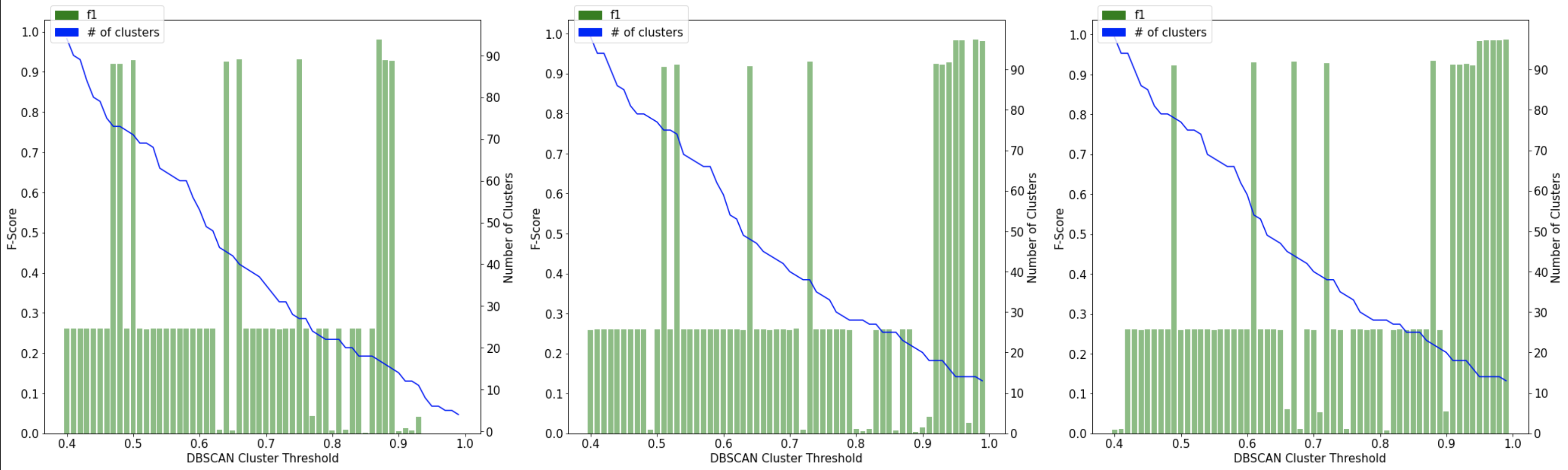}
    \label{fig:LustreResults}
    \vspace{-2em}
\end{figure*}
\begin{figure*}[t]
    \centering
    \includegraphics[width=1\linewidth]{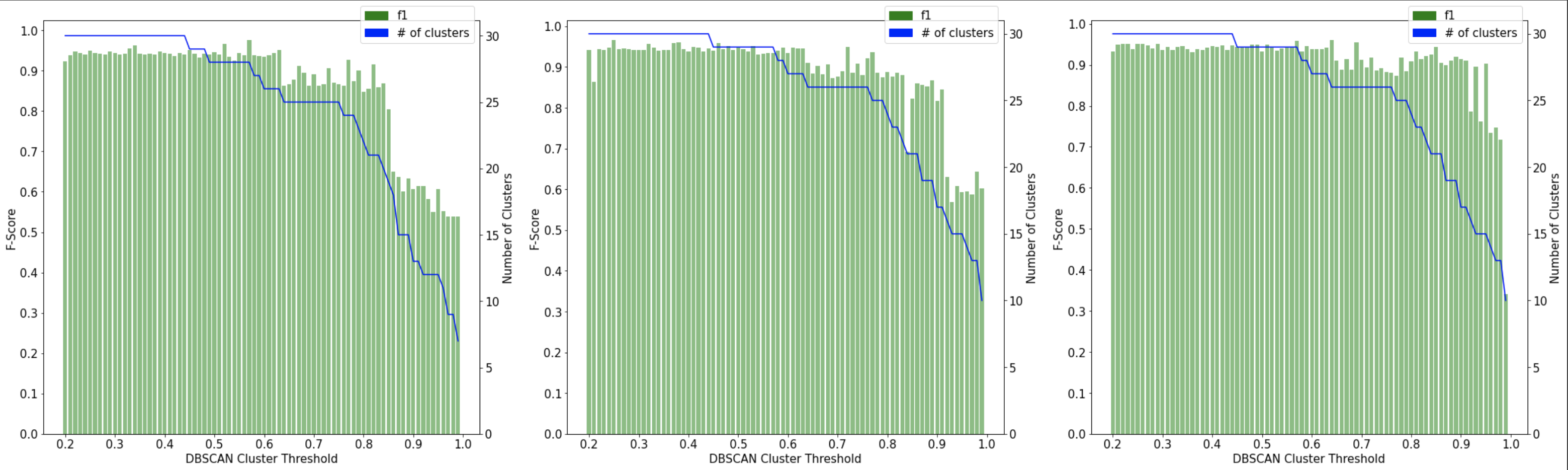}
    \caption{\small number of clusters (right $y$-axis) and F-score of the prediction (left $y$-axis) of ClusterLog under different parameter settings for \textit{Lustre} (top) and \textit{HDFS} (bottom). Adding sentiment dimension (the middle figure) shows a clear improvement in F1 scores in a larger range of DBSCAN thresholds. Also, variable number of candidates further improves the F1 scores.}
    \label{fig:HDFSResults}
\end{figure*}

\subsubsection{ClusterLog performance analysis with different settings and hyper-parameters}

The first set of results, shown in Figure~\ref{fig:HDFSResults}, represents the detailed performance of ClusterLog: 1) using numerous values for DBSCAN's epsilon parameter; 2) using sentiment dimension or not; 3) using variable or fixed number of candidates.

The leftmost graph in both of these rows shows the results of ClusterLog without the concatenation of the sentiment prediction to the semantic embedding. The middle graph shows ClusterLog's performance with the concatenation of the sentiment prediction to the semantic embedding. The right-most graph shows ClusterLog's performance with the concatenation of the sentiment prediction as well as using a variable number of candidates instead of a fixed number in the sequence model. Specifically, when the number of clusters descends below 27, the number of candidates used in the sequence model is set to the floor division of the number of clusters by 3 (eg. for 20 clusters, candidates would be set to 6). This feature was added to ClusterLog based on the intuition that when the amount of clusters becomes very low, using a relatively large amount of candidates may force the sequence model to consider a key of very low calculated probability to be normal, increasing the amount of false negatives. 

From these results, we can clearly observe not only how ClusterLog can achieve its best results by reducing a large amount of noise with high clustering thresholds, but also that the concatenation of sentiment as well as the adaptation of candidates to account for a low cluster count add to ClusterLog's ability to produce strong and consistent results.

\begin{figure}[h]
    \centering
    \includegraphics[width=0.9\linewidth]{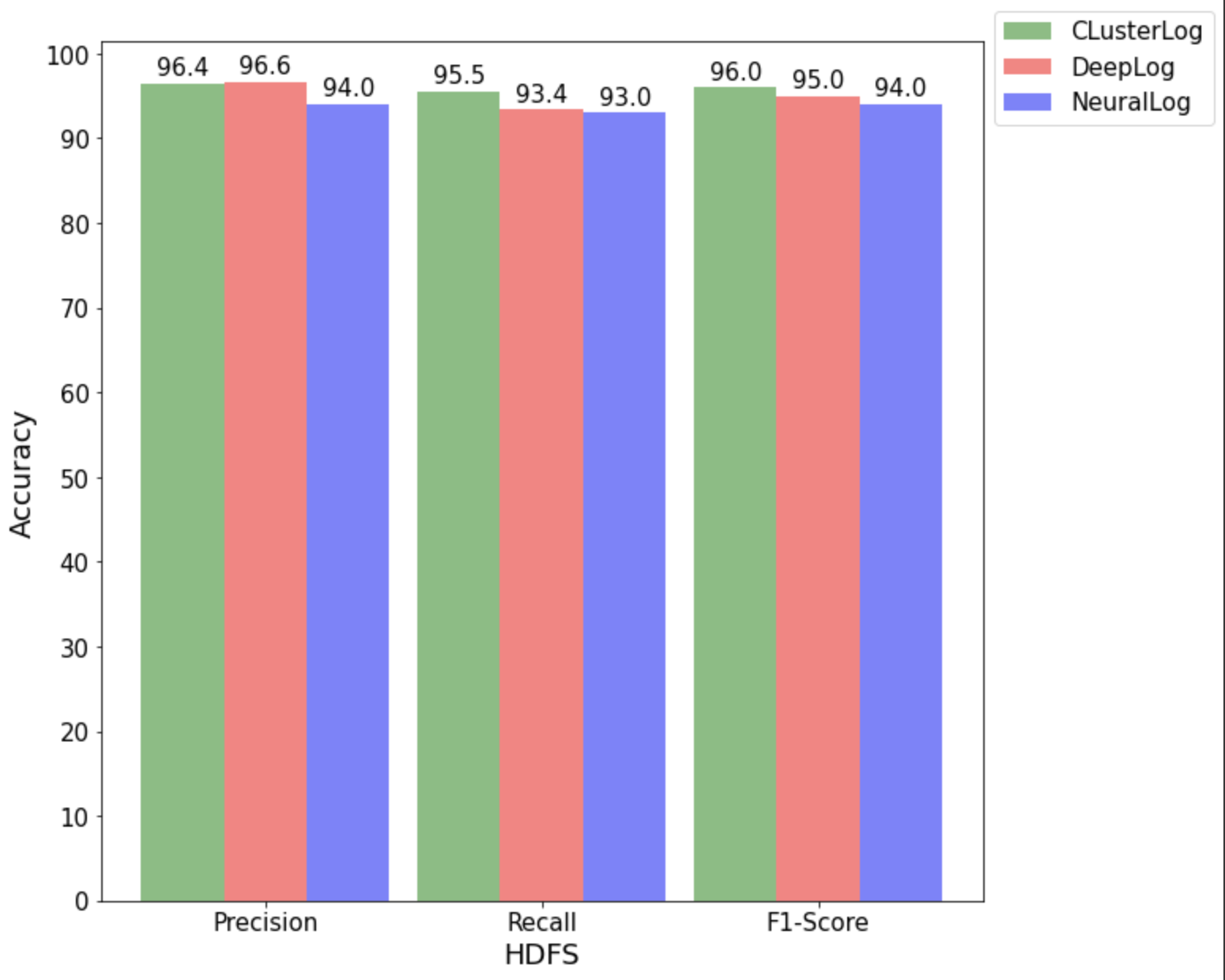}
    \caption{\small Performance comparison of ClusterLog, DeepLog, and NeuralLog on \textit{HDFS} dataset}
    \label{fig:CompareHDFS}
\end{figure}
\begin{figure}[h]
    \centering
    \includegraphics[width=0.9\linewidth]{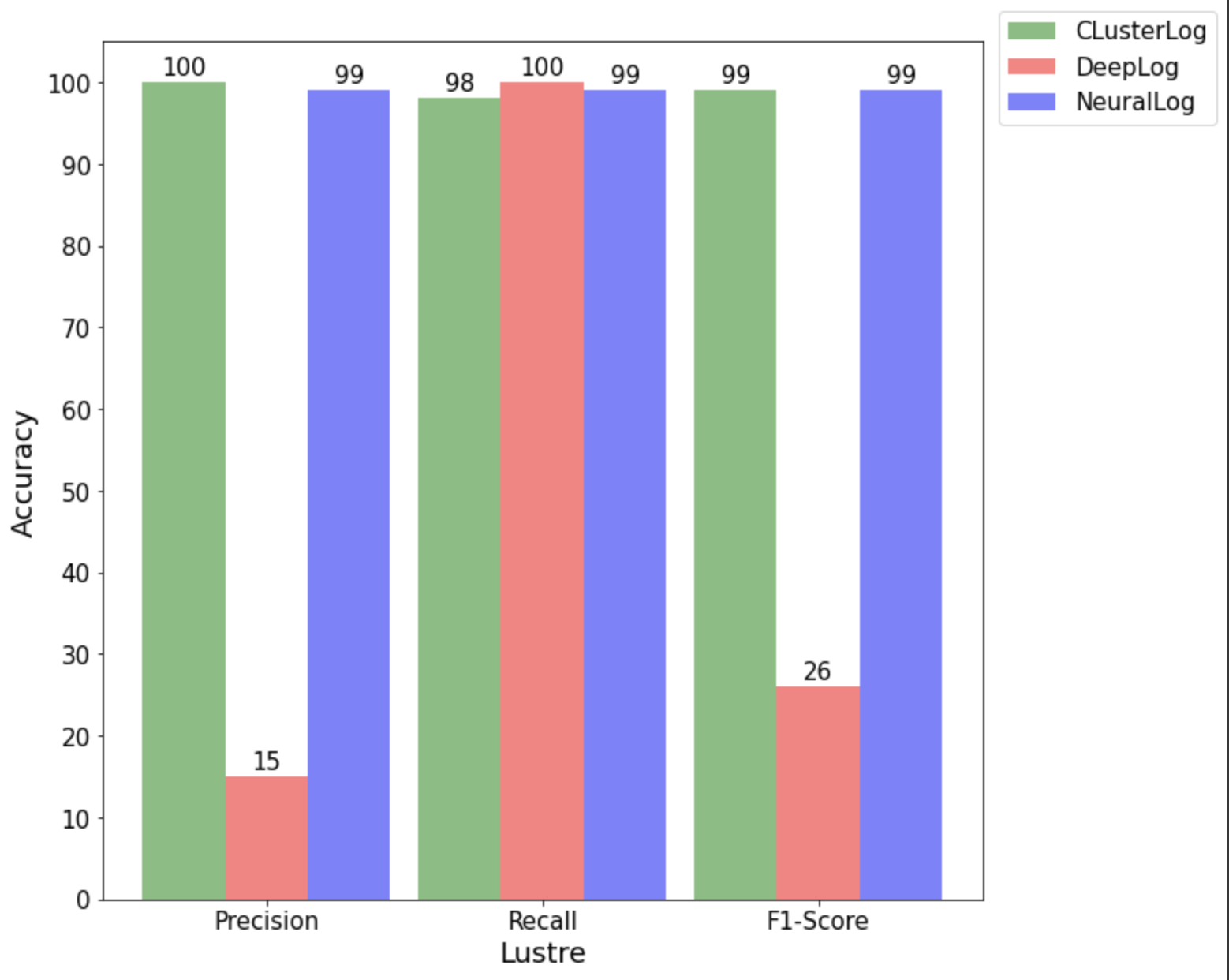}
    \caption{\small Performance comparison of ClusterLog, DeepLog and Neuralog on \textit{Lustre} dataset}
    \label{fig:CompareLustre}
\end{figure}

\subsubsection{Overall Comparison with DeepLog and NeuralLog}
In order to provide a reference among state-of-the-art anomaly detection solutions, we compared ClusterLog’s results with the results from DeepLog using the most similar training and testing setup possible. For Lustre, a sliding window of size 10 was used to predict a probability distribution among the 73 unique log key templates where a succeeding log key within the top 9 of predictions was considered normal and anything else was considered anomalous. For HDFS, the setup was the same, but because HDFS logs can be grouped into sessions using their block IDs, each session was individually used as a sequence, and the prediction of an anomaly within a session classified the entire session as anomalous.

The results shown in Figure \ref{fig:CompareHDFS} and \ref{fig:CompareLustre} for HDFS and Lustre respectively, show ClusterLog's improvement upon previous SOTA results. In the case of HDFS, while DeepLog maintains a slightly higher Precision, ClusterLog boasts a more significant improvement in Recall, primarily because clustering the most similar logs on the HDFS dataset allows ClusterLog to more easily detect what is truly a normal sequence amidst some noise. {\color{black} NeuralLog also has high precision and recall, but both of these values seem to be slightly outshined by ClusterLog and DeepLog on this train and test split.} {\color{black}Overall, ClusterLog's very high precision and high recall boost its combined F1 score slightly higher than Deeplog's and NeuralLog's}. However, due to the ability to group HDFS logs into sessions by their block ID, the nature of these sessions is already fairly low in noise. This means the application of ClusterLog will still be effective, as is shown, but the difference in performance will likely not be as large as it may be for more noisy systems. Because Lustre is a good example of one such noisy system, the comparison of results on this dataset provides evidence of this larger performance gap. 

As shown by Figure \ref{fig:CompareLustre}, the disparity in precision between DeepLog and ClusterLog is very large while it seems both approaches have a very high recall. This is because DeepLog is not able to learn the more noisy sequence effectively, and as a result classifies the vast majority of normal and abnormal logs as anomalous. {\color{black} NeuralLog does a much better job of holding up against this dataset as its vector representations for logs are more robust than DeepLog's log indexes for noisy data, but as shown later, it takes much more labeled data to extract good performance from NeuralLog due to a very large set of trainable weights.} ClusterLog was able to learn the sequence and accurately discern between normal and abnormal logs. In total, the results on both of these datasets in comparison to modern approaches verify ClusterLog's ability to generalize across different file systems and achieve strong results.

\begin{figure}[h]
    \centering
    \includegraphics[width=0.9\linewidth]{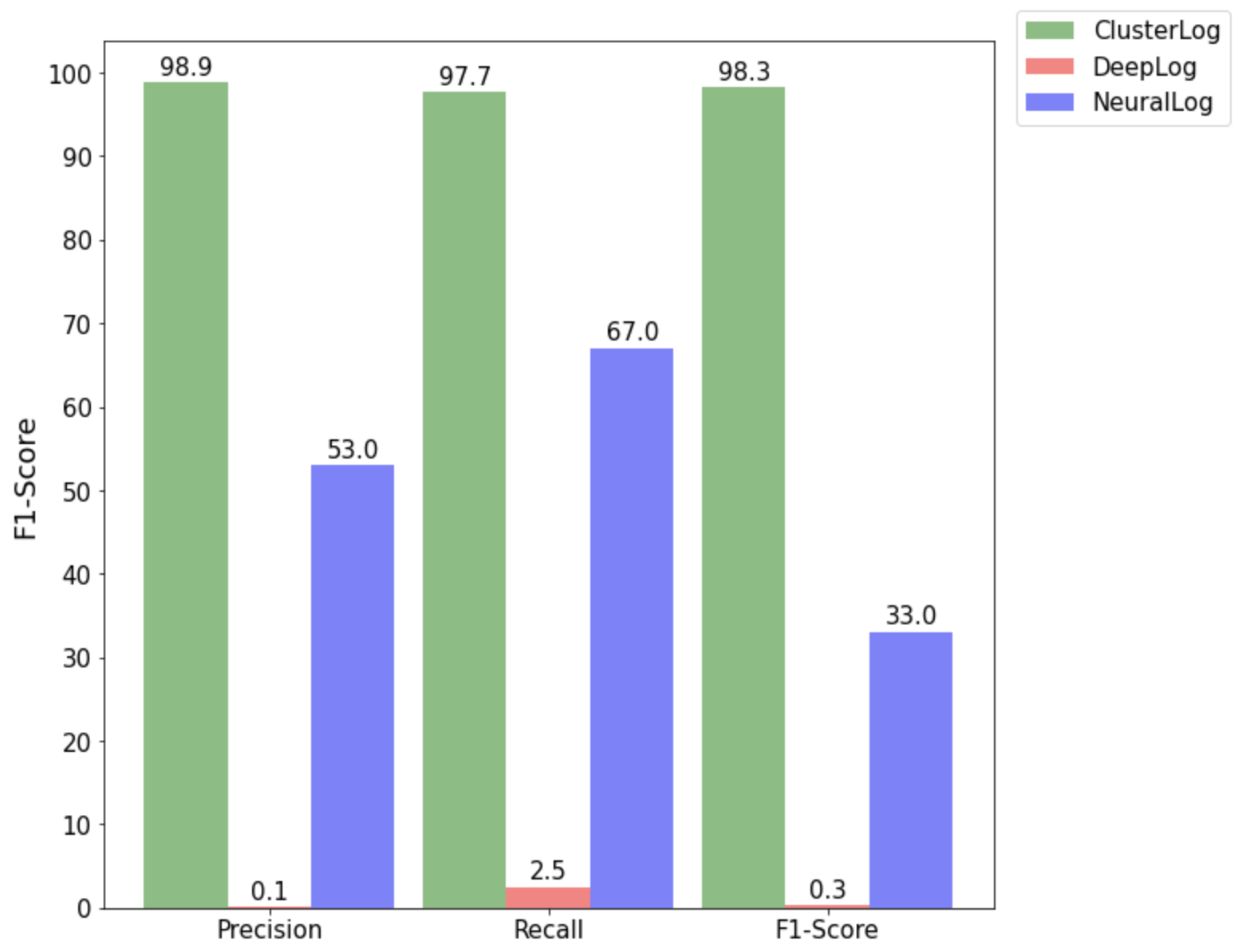}
    \caption{\small Performance comparison between ClusterLog, DeepLog, and NeuralLog using smaller training data. ClusterLog is much more stable even with less training data, showing the effectiveness of clustering logs.}
    \label{fig:Proportions}
\end{figure}

\subsubsection{The impact of training dataset size using ClusterLog}
The final comparison for ClusterLog represents the impact of using a small training set on the ability of ClusterLog, DeepLog and NeuralLog to accurately predict anomalies on the more noisy Lustre dataset. This comparison was done on the Lustre dataset because the difference in the number of clusters used by ClusterLog and DeepLog on this dataset is large, so the impact of changing the training set size should be emphasized by this fact. {As shown in Figure \ref{fig:Proportions}, the prediction results for DeepLog and NeuralLog are massively impacted as the proportion of the training set used shrinks}. {\color{black}The alternative approaches' low performance on such a small dataset can be explained by the fact that the normal interactions between a higher granularity of keys are unlikely to be explained by such a small dataset}. In contrast, the figure shows that ClusterLog can maintain the same level of results even having trained on just 1 percent of the training set (1504 logs). By reducing the number of keys through clustering, ClusterLog is much more likely to not only see all of the keys enough times but also learn the normal interactions between them, ultimately leading to much stronger results on small datasets.

%% file: RelatedWork.tex
{\color{black}There is a large catalogue of both preprocessing and anomaly detection approaches for file system logs in existence today~\cite{vaarandi_logcluster_2015, vaarandi_data_2003, nagappan_abstracting_2010, fu_execution_2009, tang_logsig_2011, shima_length_2016, mizutani_incremental_2013, zhu_tools_2019, cinque2012event, roy2015perfaugur, hansen1993automated, oprea2015detection, yamanishi2005dynamic, debnath2018loglens, chen2004failure, liang2007failure, du2017deeplog, meng2019loganomaly, xu2009online, xu2009detecting, lou2010mining, lin2016log, das2018doomsday}. Among preprocessing techniques, the primary approaches include frequent pattern mining, clustering, and heuristic methods. Frequent pattern mining approaches extract frequently occurring patterns, with the ultimate goal of grouping the logs which have the highest degree of similarity ~\cite{vaarandi_data_2003, vaarandi_logcluster_2015, nagappan_abstracting_2010}. The clustering-based preprocessing techniques employ clustering algorithms to group similar logs. The individual approaches in this category generally differ in their approach towards measuring similarity, as some may employ weighted edit differences on log pairs~\cite{fu_execution_2009}, while others may fit a given set of raw logs to a predefined set of events~\cite{tang_logsig_2011}. The final category of preproccessing approaches is heuristic-based approaches. Among these, Drain~\cite{he_drain_2017} is the most popular and relies on a fixed depth tree to extract log templates.} ClusterLog belongs to the clustering-based method with new designs using semantic and sentimental features of logs.

Anomaly detection approaches can be categorized into non-machine learning and machine learning approaches. 
Of the more successful non-machine learning approaches, the rule-based approaches are well seen, where domain experts formulate a set of rules which can be used to classify logs as anomalies ~\cite{cinque2012event,roy2015perfaugur, hansen1993automated,oprea2015detection,debnath2018loglens}. 
Among machine learning techniques, pattern-seeking \cite{xu2009online, xu2009detecting} and invariant mining \cite{lou2010mining} based algorithms have proven effective on a variety of file system logs, but their results do not hold up on the irregularity of logs which do not include session identifiers~\cite{zhang2021sentilog}. 
Additionally, These approaches do not hold up to more recent deep learning-based solutions, 
which learn sequences of logs using deep neural networks. The first approach, LogAnomaly\cite{meng2019loganomaly}, borrows from NLP by proposing a log key embedding technique which vectorizes log keys based on synonyms and antonyms and calculates similarity between vector embeddings to predict anomalies. An additional approach in this domain, LogBERT\cite{GuoLogBert}, utilizes the Bidirectional Encoders Represented by Transformers model which has provided STOA Results in multiple domains. In this approach, it is not the next log key that is predicted, but rather a given sequence that is masked and then classified as normal or abnormal. More recently, similar attention-based approaches have shown very strong results across a variety of Distributed File Systems. Among these approaches, LogRobust\cite{zhang2019robust} and Nueralog\cite{le2021log} show the most promising results. While the encapsulated models are slightly different, both of these approaches utilize attention mechanisms to classify a sequence of logs represented by content embedding vectors. These approaches lend well to unseen logs. However, training accurate classification on such high dimensional embedding input requires a large amount of labeled training data. Additionally, the lack of grouping characteristics such as sequence identifiers in some logs and the resultant irregularity makes it difficult for these solutions to be effectively applied. ClusterLog is proposed to better work with these advanced models to achieve better performance and higher training efficiency.

%% file: Conclusion.tex
In this work, through ClusterLog, we have shown that granularity reduction in log key sequences is a viable approach to improve log anomaly detection in general. A key area where this method of anomaly detection outperforms previous approaches is in non-session parallel file system logs, which are highly irregular and noisy. Additionally, this work shows that granularity reduction in log sequences may allow users to reduce their effort by allowing for more lenient pre-processing, as well as smaller labeled datasets. This all is because of the effective reduction of noise and retention of important sequence information achieved by a good clustered representation of the file system log keys.

In the future, there are two primary areas in this work which are intended to be improved and many others which can be explored. To begin, due to ClusterLog’s high dependence on the accuracy of sentence and sentiment embedding models, improvements in both of these areas could prove to be of value. 
The first direction of improvement is further exploration of clustering techniques to find an even more clear and easy-to-apply way of selecting the right clustering hyperparameters. 
The second direction is to further provide evidence for CLusterLog’s generalizability by applying it to a larger scope of log systems. In addition to more parallel and distributed file systems, ClusterLog should prove its viability against other systems such as Kubernetes.